\documentclass[pra,aps,epsf,nofootinbib,
notitlepage,showpacs,10pt]{revtex4-1}

\usepackage{color}
\usepackage{float}
\usepackage{graphicx}
\usepackage{amsmath}
\newcommand{\bracket}[2]{\left<#1|#2\right>}
\newcommand{\bfly}[2]{\ket{#1}\!\!\bra{#2}}
\newcommand{\eq}[1]{Eq.~(\ref{#1})}
\newcommand{\se}[1]{Sec.~\ref{#1}}
\newcommand{\avg}[3]{\left<#1|#2|#3\right>}

\input{Qcircuit} 
\usepackage{graphicx} 
\graphicspath{ {rrImages/} } 
\usepackage{tikz} 
\usetikzlibrary{matrix,positioning}
\usepackage{amsfonts}
\usepackage{amssymb}

\begin{document}

\markboth{Rebekah Herrman and Travis S.~Humble$^{1}$}
{Continuous-time Quantum Walks on Dynamic Graphs}

\title{Continuous-Time Quantum Walks on Dynamic Graphs}
\author{Rebekah Herrman}
\affiliation{Department of Mathematical Sciences\\University of Memphis\\Memphis, Tennessee USA 38152-3520\\email: rherrman@memphis.edu}
\author{Travis S.~Humble}
\affiliation{Quantum Computing Institute\\Oak Ridge National Laboratory\\Oak Ridge, Tennessee USA 37831-6015\\email: humblets@ornl.gov}

\begin{abstract}
Continuous-time quantum walks (CTQWs) on static graphs provide efficient methods for search and sampling as well as a model for universal quantum computation. We consider an extension of CTQWs to the case of dynamic graphs, in which an ordered sequence of graphs governs free evolution of the quantum walk. We then consider how perfect state transfer during the quantum walk can be used to design dynamic graphs that implement a universal set of quantum logic gates. We give explicit examples for a complete logical basis, and we validate implementations using numerical simulations for quantum teleportation and addition circuits. Finally, we discuss the potential for realizing CTQWs on dynamic graphs using actively controlled quantum optical waveguides.
\end{abstract}
\thanks{This manuscript has been authored in part by UT-Battelle, LLC, under contract DE-AC05-00OR22725 with the US Department of Energy (DOE). The US government retains and the publisher, by accepting the article for publication, acknowledges that the US government retains a nonexclusive, paid-up, irrevocable, worldwide license to publish or reproduce the published form of this manuscript, or allow others to do so, for US government purposes. DOE will provide public access to these results of federally sponsored research in accordance with the DOE Public Access Plan (http://energy.gov/downloads/doe-public-access-plan)}

\maketitle

\section{Introduction}
Quantum walks offer a unique paradigm for using quantum mechanics to perform computation \cite{aharonov1993quantum}, where a walk may represent either the discrete or continuous-time propagation of a quantum state over a graph \cite{kempe2003quantum,Venegas-Andraca2012}.
In a continuous-time quantum walk (CTQW), free evolution of an $N$-dimensional quantum state under a Hamiltonian is represented by probability amplitudes assigned to each vertex in a graph on $N$ vertices. The CTQW was originally envisioned as a method for sampling decision trees \cite{Farhi1998} and was later applied to a variety of search and sampling problems on $d-$ dimensional lattices, searches on balanced trees, as well as quantum navigation of networks \cite{aaronson2003quantum,childs2004spatial,gamble2010two,sanchez2012quantum,philipp2016continuous,li2017renormalization}. Moreover, Childs has shown that CTQWs on time-independent graphs offer a novel model for universal quantum computation \cite{Childs2009,Childs2013}, while Qiang et al.~have described how efficient implementations of CTQWs may be useful for comparing the broader computational power of quantum computing to conventional computing models~\cite{qiang2016efficient}.
\par
In a typical CTQW, the Hamiltonian is interpreted as the connectivity of the underlying graph on which the quantum state evolves.  The graph connectivity determines the evolution of the quantum state and specific graphs have been found to demonstrate well-defined quantum walk behaviors. For example, perfect state transfer occurs in a quantum walk when the amplitude assigned to a subset of vertices transfers with unit probability to a distinct vertex set within a well-defined period of evolution \cite{kay2010perfect}. Kendon and Tamon have surveyed perfect state transfer for a number of several specific graphs including the singleton graph, $K_1$, the complete graph on two vertices, $K_2$, the path graph on three vertices, and the cycle on four vertices, $C_4$ \cite{Kendon2011}. Perfect state transfer has also been shown to exist for graphs on more vertices, including certain graph products, weighted join graphs, and quotient graphs \cite{Coutinho2016, AngelesCanul2010, Bachman2012}.
\par
The versatility of CTQWs across many known types of graphs motivates our consideration for how quantum walks may behave on dynamic graphs. We define a dynamic graph as a well-defined sequence of static graphs in which the CTQW evolution changes at specific transition times. In the dynamic graphs discussed below, we use perfect state transfer under the component static graphs to demonstrate how more complex unitary processes can be realized. We provide explicit realizations of quantum walks on dynamic graphs for realizing a complete set of computational gates, and we then illustrate how compositions of multiple walks correspond to examples of quantum circuits. This formalism establishes a connection between CTQWs on dynamic graphs and the gates found in the conventional quantum circuit model. 
\par
Our approach to quantum walks on dynamic graphs shares similarities with Childs' model for universal quantum computation \cite{Childs2009,Childs2013}. Both approaches draw on the use of unweighted and relatively sparse graphs to formalize state transfer as well as the composition of such graphs to describe more complex operations. However, the models differ in the types of underlying graphs as Childs relies on strictly static graphs while we employ dynamic graphs. Another closely related model is the hybrid quantum walk proposed by Underwood and Feder, which combines concepts from both continuous and discrete walk models \cite{Underwood2010}. In that work, a series of weighted adjacency matrices corresponding to distinct graphs are used to propagate a quantum state. They refer to this model as a discontinuous quantum walk, where free evolution is again based on widgets that control propagation dynamics. Underwood and Feder emphasize the use of a dual-railing encoding to represent individual qubits and the interleaving of continuous and discrete quantum walks to perform computation. By comparison, we design quantum walks on dynamic graphs to implement a sequence of continuous-time evolutions that perform quantum logic using perfect state transfer in the native vertex space. Du et al.~considered the task of designing a quantum walk to implement a single-qubit $X$ gate by walking on a single static, weighted graph \cite{Du2018}, whereas our work develops implementations for a complete gate set using dynamics graphs. Chakraborty et al.~have explored spatial search using CTQW on time-ordered sequences of random graphs, for which they demonstrated a threshold for the optimal run time using Grover's algorithm \cite{Chakraborty2017}, while our work uses deterministic, time-ordered sequences to carry out discrete logic gates. 
\par
The paper is organized as follows. Following a review of CTQWs on static graphs in \se{sec:ctqw},  we describe quantum walks on disconnected graphs in \se{sec:qwdg} and dynamic graphs in \se{sec:qwtg}. Using this formalism, we design a series of quantum walks that implement elementary logic gates in \se{sec:qweg}, and we demonstrate how the dynamic graphs may be composed to correspond with gate-based circuits in \se{sec:qwc}. We offer a discussion on these results in \se{sec:con}, where we establish a connection between dynamic quantum walks and current approaches to designing quantum computing hardware based on optical waveguides.
\section{Continuous-time Quantum Walks}
\label{sec:ctqw}
Consider an undirected graph $G = (V, E)$ with a canonically labeled vertex set $V = \{0, 1, \ldots, N-1\}$ of $N$ vertices and an edge set $E = \{(i,j): i\sim j\}$. We allow no multi-edges in the graph, i.e., there can be at most one edge incident with any two vertices. However, we do allow for a self loop on a vertex $ v \in V$ if and only if there does not exist  $u \in V$ such that $u \neq v$ and $u\sim v$. Additionally, the edges of $G$ are undirected. 
Let $B_G = \{\ket{j}: \forall j \in V\}$ be a linearly independent basis for the complex vector space $\mathbb{C}^{N}$ with the inner product $\bracket{j}{k} = \delta_{jk}$. Graphs $G$ and $G^\prime = (V^\prime, E^\prime)$ have the same basis if $V = V^\prime$. The Hamiltonian for the graph $G$ is denoted as $H_G$ and is defined as the adjacency matrix of the graph as given by the edge set $E$. The adjacency matrix $A$ of $G$ is a 0-1 valued $N \times N$ matrix such that for $u,v \in V(G)$, if $u\sim v$, $A_{u,v}=A_{v,u}=1$, and $0$ otherwise. We will use the convention that if a vertex $v$ is not adjacent to any other vertices then $A_{v,v}=1$, a convention also used in studies of classical random walks. The resulting real-valued adjacency matrix $A$ is symmetric about the main diagonal.
\par
We define the quantum state of a graph $G$, or graph state for short, as a normalized vector $\ket{\psi_G} \in B_{G}$ such that
\begin{equation}
\ket{\psi_G} = \sum_{j\in V}{c_j \ket{j}}
\end{equation}
with $c_{j} \in \mathbb{C}$ and
\begin{equation}
\bracket{\psi_G}{\psi_G} = \sum_{j\in V}{|c_{j}|^2} = 1
\end{equation}
For a continuous-time quantum walk, the graph state transforms with respect to time $\tau$ under the Schr{\"o}dinger equation
\begin{equation}
\label{eq:tdse}
i\hbar\frac{\partial \ket{\psi_G (\tau)}}{\partial \tau} = H_G\ket{\psi_G (\tau)}
\end{equation}
where $\hbar$ is Planck's constant divided by $2\pi$. When the Hamiltonian is constant over the interval $[t_0,t]$, the formal solution to \eq{eq:tdse} is given by the propagation operator
\begin{equation}
\label{eq:ug}
U_G(t,t_0) = e^{-i H_G (t-t_0)/\hbar}
\end{equation}
such that
\begin{equation}
\ket{\psi_{G}(t)} = U_G(t,t_0)\ket{\psi_{G}(t_0)}
\end{equation}
where the boundary condition $\ket{\psi_{G}(t_0)}$ is the state at time $t_0$. The propagation operator $U_G$ is unitary since $H_G$ is Hermitian. We say that a graph $G$ admits perfect state transfer between unique vertices $u,v \in V(G)$ at time $t$ if
\begin{equation}
U_G(t,0)\ket{u} = a \ket{v}
\end{equation}
where $a \in \mathcal{C}$ such that $ |a|=1$.
\par
There are several well-known examples that illustrate perfect state transfer using CTQW on static graphs. The singleton graph $K_1$ has vertex set $V = \{0\}$ and an empty edge set $E =\varnothing$. As the lone vertex $\ket{0}$ is adjacent to no other vertices during the CTQW, we represent the unitary dynamics by a self-loop. The $K_1$ Hamiltonian is then represented in its eigenbasis as 
\begin{equation}
H_{K_1} = \lambda_1 \bfly{0}{0},
\end{equation}
where $\lambda_1$ is the real-valued energy eigenvalue, and the normalized state
\begin{equation}
\ket{\psi_{K_1}(t_0)} = c_0(t_0) \ket{0}
\end{equation}
has $|c_0| = 1$ for all time such that
\begin{equation}
\label{eq:psik1}
\ket{\psi_{K_1}(t)} = e^{-i \nu_{1} t} \ket{0}
\end{equation}
where $\nu_{1} = \lambda_1/\hbar $ is the frequency. 
\par
\begin{figure}
\centering
\includegraphics{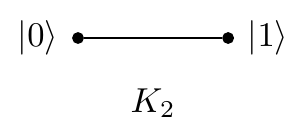}
\caption{The $K_2$ graph supports perfect state transfer between the two vertices labeled by the single-qubit computational basis states $\ket{0}$ and $\ket{1}$.}
\label{fig:k2}
\end{figure}
\par
As a second example, the complete graph on two vertices $K_2$ shown in Fig.~\ref{fig:k2} has vertex set $V = \{0,1\}$ and edge set $E= \{(0,1)\}$. We specify the Hamiltonian for $K_2$ as the free evolution operator over degenerate basis states, i.e., $\avg{0}{H_{K_2}}{0} = \avg{1}{H_{K_2}}{1}$, which offers a natural representation of a qubit in a degenerate eigenbasis. Setting this eigenenergy to zero, the Hamiltonian is represented as
\begin{equation}
H_{K_2} = \lambda_2 \left(\bfly{0}{1}+\bfly{1}{0}\right)
\end{equation}
where the eigenvalue $\lambda_2$ defines the energy scale  and the characteristic frequency $\nu_2 = \lambda_2 /\hbar$. 
The time propagator for $K_2$ may be decomposed by series expansion as 
\begin{equation}
\label{eq:uk2}
U_{K_2}(t,t_0) = \cos[\nu_2 (t-t_0)] \mathbb{I}_2 - i \sin[\nu_2 (t-t_0)] H_{K_2}
\end{equation}
where $\mathbb{I}_{N}$ is the $N$-dimensional identity matrix. The $K_2$-graph state evolves as
\begin{equation}
\begin{array}{rcl}
\ket{\psi_{K_2}(t)} &=& \left(c_0 \cos[\nu_2 (t-t_0)] - i c_1 \sin[\nu_2 (t-t_0)] \right) \ket{0} + \left(c_1 \cos[\nu_2 (t-t_0)] - i c_0 \sin[\nu_2 (t-t_0)] \right) \ket{1}
\end{array}
\end{equation}
which is capable of perfect state transfer up to a trivial phase factor for propagation time $t = \frac{\pi}{2 \nu_2}$ \cite{Kendon2011}. 
\par
\begin{figure}[h]
\includegraphics{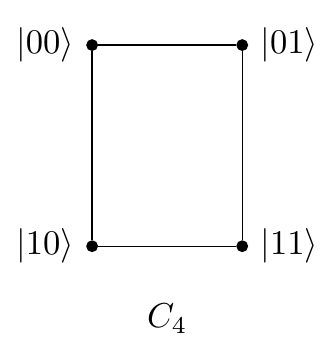}
\caption{The $C_4$ graph supports perfect state transfer between vertices labeled by the two-qubit computational states.}
\label{fig:c4}
\end{figure}

As a final example, the cycle graph $C_4$ shown in Fig.~\ref{fig:c4} has a vertex set $V = \{0, 1, 2, 3\}$, edge set $E = \{ (0,1), (0,2), (1,3), (2,3) \}$ and Hamiltonian
\begin{equation}
H_{C_4} = \lambda_4 \left(\bfly{0}{1} + \bfly{0}{2} + \textrm{H.C.}\right)
\end{equation}
where $\textrm{H.C.}$ denotes the Hermitian conjugate, $\lambda_4$ is the energy scale, and $\nu_4 = \lambda_4/\hbar$ defines the characteristic frequency. The propagation operator may be decomposed as
\begin{equation}
\begin{array}{lcr}
U_{C_4}(t,t_0) &=& \mathbb{I}_4+\frac{1}{2}\cos({2\nu_4 (t-t_0)}) H_{C_4}^2 - \frac{i}{2}\sin({2 \nu_4(t-t_0)}) H_{C_4}
\end{array}
\end{equation}
to yield the state $\ket{\psi_{C_4}(t)}$ with coefficients in the nodal basis as
\begin{equation}
\begin{array}{rcl}
c_{0}(t) &=&  \frac{1}{2} \left[c_0\left(1+\cos \left(2 t\right)\right)-i\sin \left(2t\right)\ \left(c_1+c_2\right)  +c_3 \left(-1+\cos \left(2t\right) \right)\right]
\end{array}
\end{equation}
\begin{equation}
\begin{array}{rcl}
c_{1}(t) &=&  \frac{1}{2} \left[-i\sin \left(2t\right)\ \left(c_0+c_3\right)+c_1\left(1+\cos \left(2 t\right)\right) +c_2 \left(-1+\cos \left(2t\right) \right)\right]
\end{array}
\end{equation}
\begin{equation}
\begin{array}{rcl}
c_{2}(t) &=&  \frac{1}{2} \left[-i\sin \left(2t\right)\ \left(c_0+c_3\right)+c_1\left(-1+\cos \left(2 t\right)\right)  +c_2 \left(1+\cos \left(2t\right) \right)\right]
\end{array}
\end{equation}
and
\begin{equation}
\begin{array}{rcl}
c_{3}(t) &=&  \frac{1}{2} \left[c_0\left(-1+\cos \left(2 t\right)\right)-i\sin \left(2t\right)\ \left(c_1+c_2\right)   +c_3 \left(1+\cos \left(2t\right) \right)\right]
\end{array}
\end{equation}
Perfect state transfer in $C_4$ is a special instance of the case of an $N$-dimensional hypercube \cite{Kendon2011}, which has been shown by Christandl et al.~to be capable of perfect state transfer for all $N$ at time $t= \frac{\pi}{2 \nu_N}$  \cite{Christandl2004,Christandl2005}. For the remainder of our presentation, we will simplify the analysis to the case that $\nu_{N} = 1 $ for $k=1,2,4$ and we will set $\hbar=1$ for convenience.
\section{Quantum walks on Disconnected Graphs}
\label{sec:qwdg}
We now consider quantum walks on disjoint graphs $G_1$ and $G_2$ with $G_j = (V_j, E_j)$, where the disjoint union $G = G_1 + G_2$ has vertex set $V = V_1 \cup V_2$ and edge set $E = E_1 \cup E_2$.  We require that $G_1$ and $G_2$ are disconnected graphs, termed components of the graph $G$, and that $V_1 \cap V_2 = \varnothing$. The basis for the disjoint union $G$ is $B_G = B_{G_1} \oplus B_{G_2}$ and a composite quantum state for $G$ takes the form
\begin{equation}
\ket{\psi_G} = \ket{\psi_{G_1}} \otimes \ket{\psi_{G_2}}
\end{equation}
with $\oplus$ the direct sum and $\otimes$ the Kronecker product. 
The corresponding Hamiltonian is defined as $H(G) = H_{G_1} \oplus H_{G_2}$, which yields decoupled equations of motion
\begin{equation}
i\frac{\partial \ket{\psi_{G_j} (t)}}{\partial t} = H_{G_j}\ket{\psi_{G_j} (t)}\hspace{1cm}j=1,2
\end{equation}
and a composite time propagator 
\begin{equation}
U_{G_1 + G_2}(t,t_0) = e^{-i H_{G_1} (t-t_0)} \otimes e^{-i H_{G_2} (t-t_0)}
\end{equation}
The graph state of $G$ is modeled by two disconnected states $\ket{\psi_{G_1}} \in B_{G_1}$ and $\ket{\psi_{G_2}} \in B_{G_2}$ and
\begin{equation}
\ket{\psi_{G}(t)} = U_{G_1}(t,t_0)\ket{\psi_{G_1}(t_0)} \otimes U_{G_2}(t,t_0)\ket{\psi_{G_2}(t_0)}
\end{equation}
\par
As an example, consider the empty graph on $N$ vertices $\bar{K}_N$, which is the complement of the complete graph $K_N$ and expressed as the union
\begin{equation}
\bar{K}_n = \bigcup_{j=0}^{N-1}{K_1^{(j)}}
\end{equation}
where $K_1^{(j)}$ is the singleton graph with vertex label $j$. The composite Hamiltonian is the direct sum of $N$ singleton Hamiltonians,
\begin{equation}
H_{\bar{K}_n} = \bigoplus_{j=0}^{N-1}{H_{K_1^{(j)}}},
\end{equation}
and the quantum state is defined as the Kronecker product of the individual states.
\begin{equation}
\ket{\psi_{\bar{K}_{N}}} = \bigotimes_{j=0}^{N-1}{\ket{\psi_{K_{1}{^{(j)}}}}}
\end{equation}
in the basis formed from the union of these graphs
\begin{equation}
B_{\bar{K}_N} = \bigcup_{j=0}^{N-1}{B_{K_{1}^{(j)}}}
\end{equation}
The graph state for the $j$-th singleton graph is now given as
\begin{equation}
U_{K_{1}^{(j)}}(t,t_0)\ket{j} = e^{-i \nu_{1}^{(j)} t} \ket{j},
\end{equation}
with $\nu_{1}^{(j)}$ the energy eigenvalue of the $j$-th vertex. We will assume that the vertices are indistinguishable and therefore $\nu_{1}^{(j)} = \nu_1$ for all $j$. Thus, the Hamiltonian of these $N$ disjoint identical vertices
\begin{equation}
H_{\bar{K}_N} = \bigoplus_{j=0}^{N-1}{\lambda_1 \bfly{j}{j}} = \lambda_1 \mathbb{I}_N
\end{equation}
is proportional to the $N$-dimensional identity operator  $\mathbb{I}_N$ over the basis $B_{\bar{K}_{N}}$. This yields an $N$-fold Kronecker sum of $K_1$ states with the form of \eq{eq:psik1}.
\par
As a second example, consider the disjoint union $K_2 + K_1$ with Hamiltonian 
\begin{equation}
H_{K_2 + K_1} = H_{K_2} \oplus H_{K_1}
\end{equation}
represented as 
\begin{equation}
H_{K_2 + K_1} 
= \left(
\begin{array}{ccc}
0 & \lambda_2 & 0 \\
\lambda_2 & 0 & 0 \\
0 & 0 & \lambda_1
\end{array}
\right)
\end{equation} 
The composite state of this disjoint graph propagates as
\begin{equation}
\begin{array}{l}
U_{K_2+K_{1}}(t,t_0)\ket{\psi_{K_2+K_1}} =  U_{K_2}(t,t_0)\ket{\psi_{K_2}(t_0)} \oplus  U_{K_1}(t,t_0)\ket{\psi_{K_1}(t_0)}
\end{array}
\end{equation}
and may be recast as
\begin{equation}
\arraycolsep=1.4pt\def\arraystretch{2.2}
\begin{array}{rcl}
U_{K_2+K_{1}}(t,t_0)\ket{\psi_{K_2+K_1}} & = & \left[c_1\cos(\nu_2(t-t_0))  -i c_1 \sin(\nu_2(t-t_0)) \right] \ket{0} \\
 & & + \left[c_1 \cos(\nu_2(t-t_0))  -i c_0 \sin(\nu_2(t-t_0)) \right]\ket{1} \\
 & & + c_2 e^{-i\nu_1 t} \ket{2}
\end{array}
\end{equation}
\section{Quantum Walks on Dynamic Graphs}
\label{sec:qwtg}
We next consider quantum walks on dynamic graphs, in which a dynamic graph $\mathcal{G} = \{(G_{\ell}, t_\ell)\}$ is a set of graphs $G_{\ell} = (V_{\ell}, E_{\ell})$ with associated propagation times $t_{\ell} < t_{\ell+1}$ for $ \ell \in \mathbb{Z}$. 
In subsequent discussion, we will consider the case that only the edge sets change while the vertex sets stay constant, i.e., $V_{\ell} = V$, such that the bases for all $G_{\ell}$ are the same. However, the case of changing vertex sets is equally valid as this represents the growth and reduction of the underlying Hilbert space, for example, through the addition or removal of ancillary vertices. 
\par
The Hamiltonian of a dynamic graph $\mathcal{G}$ is expressed as the weighted sum
\begin{equation}
H_{\mathcal{G}} = \sum_{\ell =0}^{L-1}{H_{G_\ell} \Pi_\ell(t)},
\end{equation}
where transitions between graphs are modulated by the functions $\Pi_{\ell}(t)$. We consider the explicit case that the $\ell$-th transition function takes the form of the $\ell$-th rectangle function:
\begin{equation}
\Pi_{\ell}(t) = \left\{\begin{array}{rl}
1 & t_\ell < t < t_{\ell+1} \\
0 & \textrm{otherwise}
\end{array}
\right.
\end{equation}
with $[t_0, t_L]$ the interval over which the entire walk is defined. The dynamics is then expressed as a sequence of propagations through the series of Schr{\"o}dinger equations
\begin{equation}
i\frac{\partial \ket{\psi_{G_{\ell}}(\tau)}}{\partial \tau} = H_{G_{\ell}} \ket{\psi_{G_{\ell}}(\tau)}, \hspace{1cm}t_{\ell} < \tau < t_{\ell + 1}.
\end{equation}
As the set of discontinuities is countable, the function is still Riemann integrable, and this system of equations yields the composite propagation operator
\begin{equation}
U_{\mathcal{G}}(t_L,t_0) = \prod_{\ell=0}^{L-1}{e^{-iH_{G_\ell}(t_{\ell+1}-t_{\ell})}}
\end{equation}
which is understood to be a product ordered from right to left with increasing index. The quantum state of the dynamic graph $\mathcal{G}$ is then defined under this operator transform as
\begin{equation}
\ket{\psi_{\mathcal{G}}(t)} = U_{\mathcal{G_{\ell}}}(t_L, t_0)\ket{\psi_{\mathcal{G}}(t_0)}
\end{equation}
with initial condition $\ket{\psi_{\mathcal{G}}(t_0)} \in B_{\mathcal{G}}$ and $\bracket{\psi_{\mathcal{G}}(t_0)}{\psi_{\mathcal{G}}(t_0)} = 1$.
\par
\begin{figure}
\centering
\includegraphics[width=0.8\columnwidth]{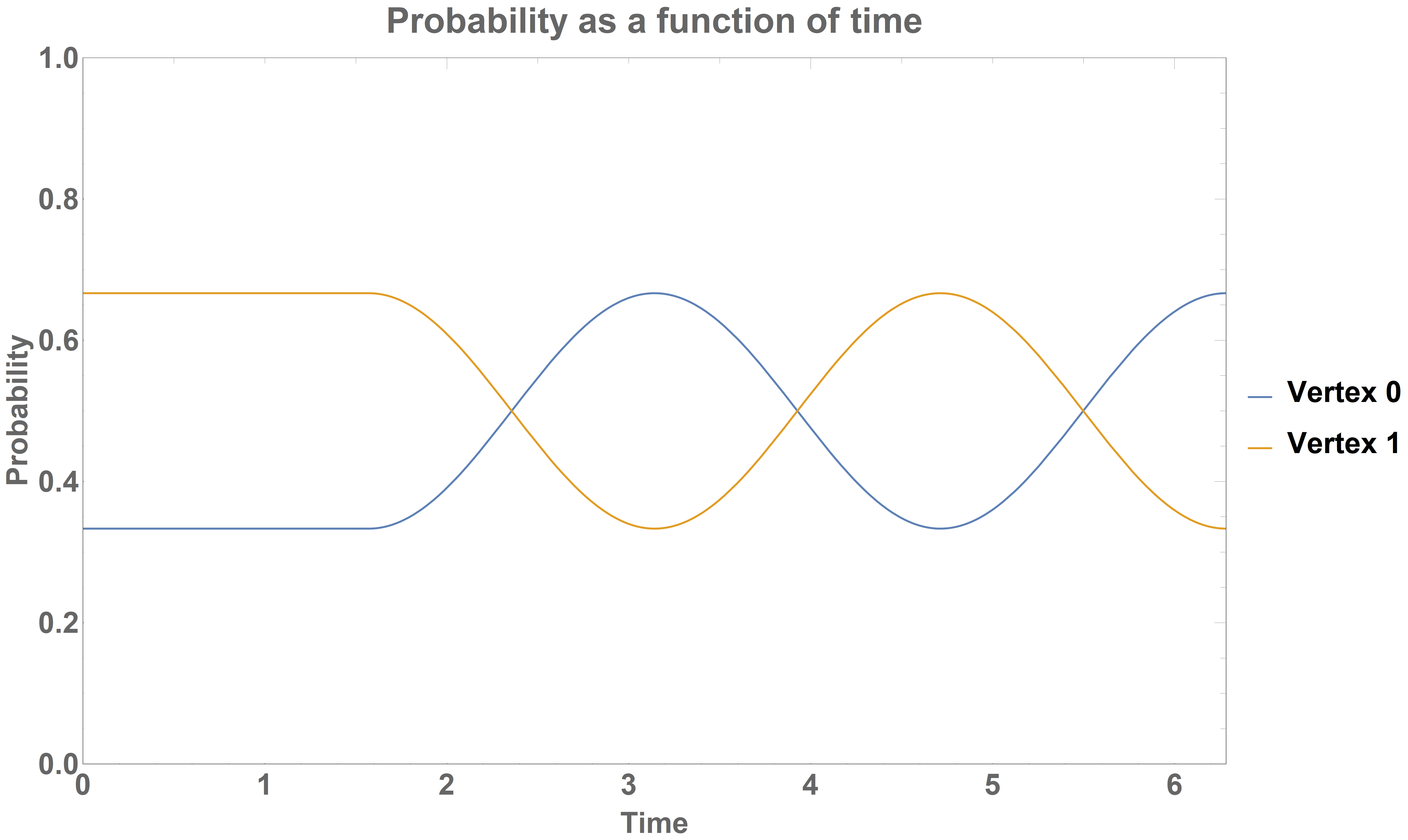}
\caption{The time-dependent probability densities of two vertices, $0$ and $1$, as the state propagates under $K_1 + K_1$ for $t=\frac{\pi}{2}$ units of time before switching to $K_2$ and propagating for an additional time $t=\frac{3\pi}{2}$. In this example, the initial state $ \sqrt{\frac{1}{3}} \ket{0} + \sqrt{\frac{2}{3}} \ket{1}$.}
\label{fig:k1k2}
\end{figure}
\begin{figure}
\centering
\includegraphics[width=0.8\columnwidth]{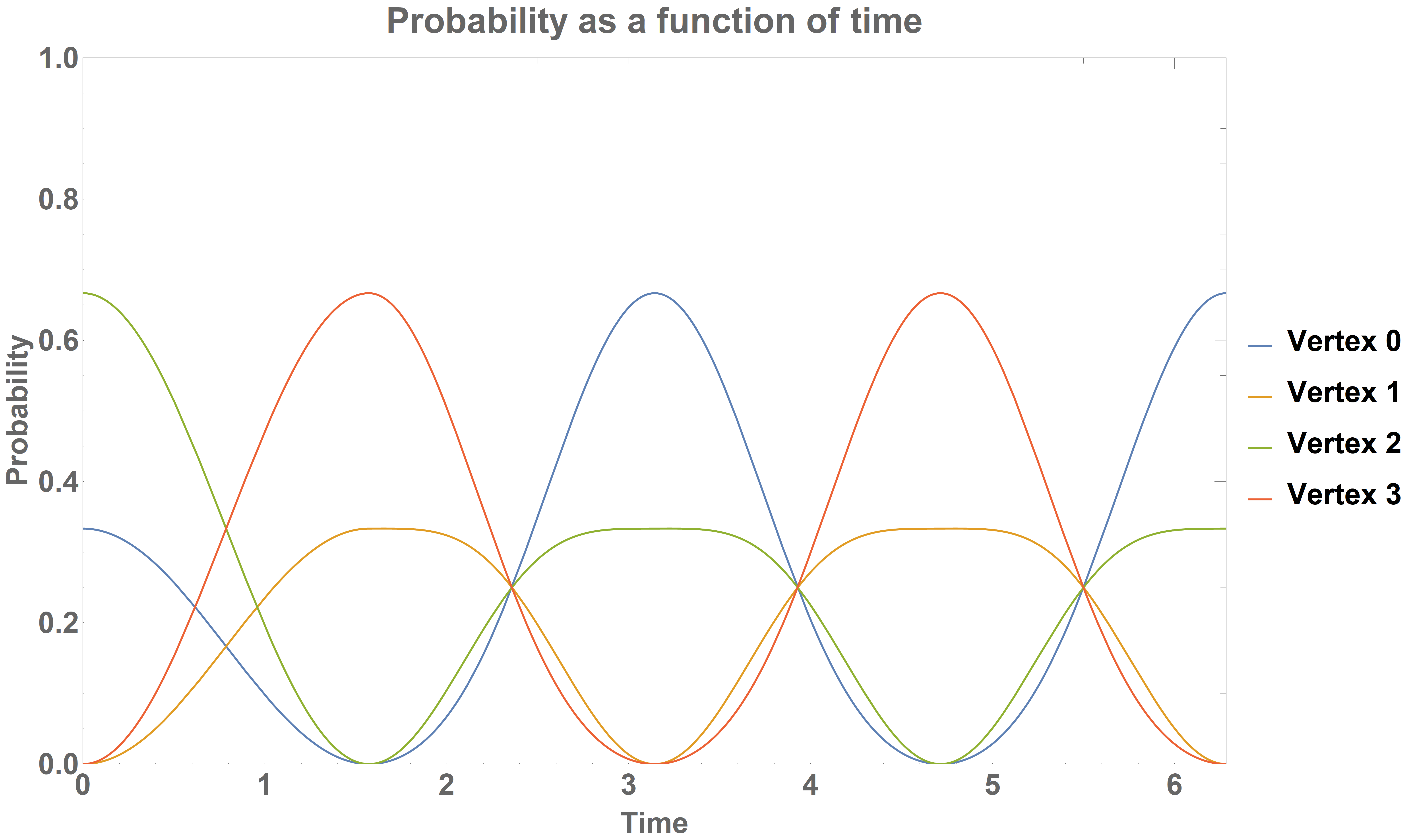}
\caption {The time-dependent probability density of four vertices, $0$, $1$, $2$, and $3$, as the state propagates under $K_2 + K_2$ for time $t = \frac{\pi}{2}$ followed by $C_4$ for time $t = \frac{3\pi}{2}$. In this example, the initial state is $ \sqrt{\frac{1}{3}} \ket{00} + \sqrt{\frac{2}{3}} \ket{10}$.}
\label{fig:K2connectC4}
\end{figure}

As a simple example of a quantum walk on a dynamic graph, consider the case of two disjoint $K_1$ graphs switched to a bipartite $K_2$. The dynamic graph is  expressed as $\mathcal{G} = \{(K_1+K_1, t_0), (K_2, t_1)\}$, where $t_0$ and $t_1$ denote the transition times. Taking the initial quantum state as a superposition over the nodal basis, Fig.~\ref{fig:k1k2} plots the time-dependent probability for each basis state with respect to the propagation time. Initially under the $K_1+K_1$, the probability remains constant until the transition time $t_1$, after which the Hamiltonian switches to $K_2$ and creates an edge between vertices. This leads to the oscillations in probability as expected by Eq.~(\ref{eq:uk2}). Figure \ref{fig:K2connectC4} is an example of two $K_2$ graphs allowed to propagate on their own and then connected as a $C_4$ and allowed to propagate again.
\section{Quantum Walks for Elementary Gates}
\label{sec:qweg}
The formalism of quantum walks on dynamic graphs may be used to realize one- and two-qubit gates within the quantum circuit model by identifying the quantum walk on a graph of $|V|=N = 2^n$ vertices with a corresponding $n$-qubit circuit. Let the vertex label $v \in V$ map to the computational basis state $\ket{v_1, v_2, \ldots, v_n}$ with $v_i$ the $i$-th coefficient in the binary expansion of the $n$-bit, non-negative integer $v$. We demonstrate several explicit examples of how few-qubit quantum gates can be realized using perfect state transfer limited to $K_1$, $K_2$, and $C_4$ graphs. We limit our CTQWs to those on $K_1$, $K_2$, and $C_4$ because the periods are all multiples of $\pi$ and achieve perfect state transfer at times $\frac{k\pi}{2}$ for $k \in \mathcal{N}$. In fact, we use the $K_1^{(i)}$ graph exclusively to add a phase factor to the $i^{th}$ qubit. We show that in some instances, such as the $Z$ gate, the realization of gate logic within the quantum walk model requires additional vertices whereas other gates, such as CNOT and CCNOT, are straightforward to realize. 
\par
The realization of elementary gates from the circuit model provides a constructive approach to demonstrate the completeness of quantum walks on dynamics graphs. While the quantum walk formalism can naturally represent any unitary of the form $\exp(iAt)$, we have imposed the restriction that the Hermitian matrix $A$ must represent the connectivity of the dynamic graph and that these graphs should be limited to a small number of vertices. By demonstrating that a complete basis of elementary gates can be constructed under these restrictions, we can then invoke the Solovay-Kitaev theorem to establish universality. The Solovay-Kitaev theorem establishes the feasibility of approximating an arbitrary unitary transformation when only a limited subset of such transformation may be accessed \cite{Dawson2006}. We demonstrate an explicit realization for a universal set of gates, including the Pauli, $H$, $T$ and CNOT gates described below, from which it follows that sequences of these gates of length $O(\log^{c}(1/\epsilon))$ may approximate an arbitrary unitary within precision $\epsilon$ for constant $c \approx 3.97$ \cite{Dawson2006}. 
\subsection{Single-qubit Gates}
\begin{figure}[H]
\centering
\includegraphics{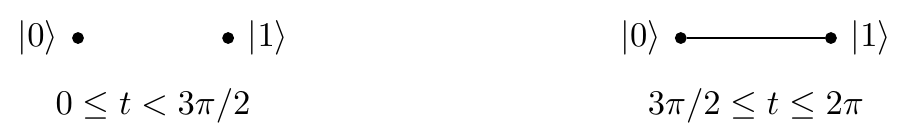}
\caption{A dynamic graph representation of the $X$ gate consists of two graphs and the associated propagation times. This sequence of CTQW executes the logical bit flip operation on the graph state.}
\label{fig:paulixgate}
\end{figure}

The Pauli gates provide a set of single-qubit operations represented in the computational basis as
\begin{align}
X = \left(
\begin{array}{cc}
0 & 1 \\
1 & 0 
\end{array}
\right),
& &
Y = \left(
\begin{array}{cc}
0 & -i \\
i & 0 
\end{array}
\right), 
& &
Z = \left(
\begin{array}{cc}
1 & 0 \\
0 & -1 
\end{array}
\right).
\end{align}
We can implement these gates exactly using perfect state transfer within a dynamic graph. For example, the $X$ gate may be implemented on two graph vertices using a quantum walk on $K_2$. For simplicity, we assuming the vertices are labeled 0 and 1 and that the graph state is initially prepared as $c_0 \ket{0} + c_1 \ket{1}$. The walk under $K_2^{(0,1)}$ for a period of $\frac{3\pi}{2}$ prepares the state $i\left(c_1 \ket{0} + c_0 \ket{1}\right)$. The resulting global phase factor of $i$ may be removed by evolving under $K_{1}^{(0)}+K_{1}^{(1)}$ for a second period of $\frac{\pi}{2}$, and we include these dynamics in our definition of the $X$ gate. The dynamic graph for the $X$ gate is defined as 
\begin{equation}
\mathcal{G}_X = \left\{\left(G_{K_2}^{(0,1)},\frac{3\pi}{2}\right), \left(G_{K_1}^{(0)}+G_{K_1}^{(1)},\frac{\pi}{2}\right)\right\},
\end{equation}
and Fig.~\ref{fig:paulixgate} provides a graphical representation. When the target pair of vertices is embedded in a larger graph state, it is understood that all other nodes evolve disjointly from the above dynamic graph. 
\par
\begin{figure}[H]
\centering
\includegraphics{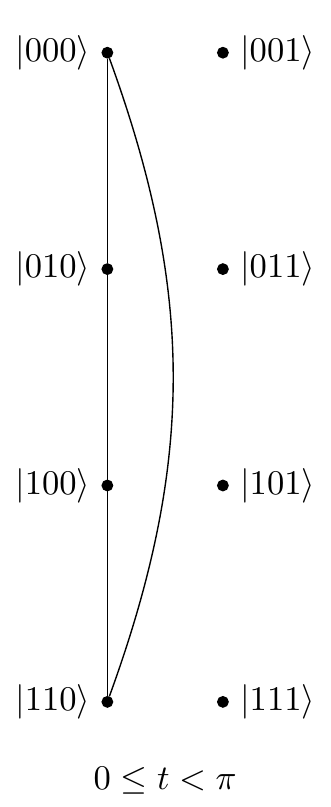}
\caption{A graphical representation of the $Z$ gate using CTQW on $\mathcal{G}_{Z}$.}
\label{fig:paulizgate}
\end{figure}

 
The $Z$ gate may be implemented using a $K_1$ and $C_4$ defined on five vertices. Notice that $\ket{001}$ must propagate as a singleton for $\pi$ units of time to flip the sign of the coefficient, however, $\ket{000}$ needs to propagate as a $C_4$ in the same time frame in order to keep its original sign. We maintain a clear correspondence with the circuit model by using a graph on eight vertices which represent the full Hilbert space for three qubits. Three of these vertices will propagate as singletons for the entirety of the walk. For example, given the initial state $c_0 \ket{0} + c_1 \ket{1}$ for a graph of $|V|=8$ vertices, the dynamic graph representing the $Z$ gate is defined as
\begin{equation}
\mathcal{G}_Z = \{(G_{C_4}^{(0,2,4,6)}+G_{K_1}^{(1)}+G_{K_1}^{(3)}+G_{K_1}^{(5)}+G_{K_1}^{(7)},\pi)\}
\end{equation}
A graphical representation of the walk for the $Z$ gate is shown in Fig.~\ref{fig:paulizgate}. Note these dynamic flips the signs of $\ket{011}, \ket{101}$, and $\ket{111}$ in addition to $\ket{001}$
\par
\par
A $Y$ gate may be derived from the commutation relations for the Pauli operators and implemented by performing the $X$ and $Z$ gates in series. An additional phase shift of $i$ is required and this may be recovered by evolving all vertices under disjoint singletons for $t=3\pi/2$. Of course, reversing the in which the $X$ and $Z$ gates are performed would change the necessary phase shift,  $-iY$. Alternatively, the $Y$ transformation may be implemented by propagating vertices $\ket{000}$ and $\ket{001}$ under $K_2$ for $\pi/2$ units of time, then allowing vertex $\ket{001}$ to propagate as a singleton for $\pi$ units while simultaneously allowing $\ket{000}$ to propagate as a $C_4$ to three new vertices. The dynamic graph for the latter $Y$ operation is given as
\begin{equation}
\mathcal{G}_Y = \{\left(G_{K_2}^{(0,1)}+G_{K_1}^{(2)}+G_{K_1}^{(3)}+G_{K_1}^{(4)},\frac{\pi}{2} \right), \{\left(G_{K_1}^{(1)}+G_{C_4}^{(0,2,3,4)},\pi \right)\ \}, 
\end{equation}
\begin{figure}
	\includegraphics{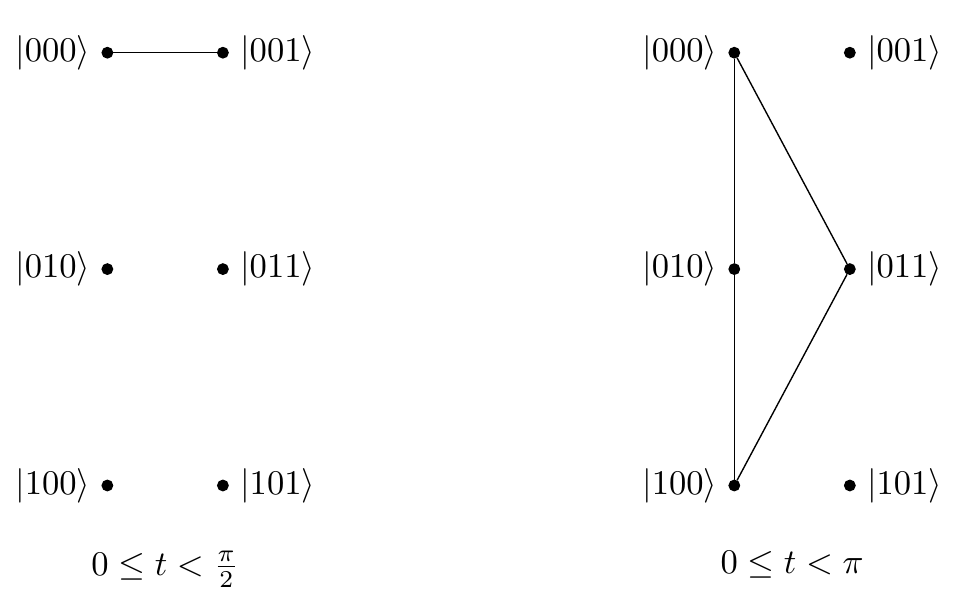}
	\caption{A graphical representation of the $Y$ gate using CTQW on $\mathcal{G}_{Y}$.}
	\label{fig:pauliygate}
\end{figure}
Completing the Pauli group, we note that the identity gate may be implemented using a number of different dynamic graphs. This includes assigning every vertex to propagate under the singleton graph for $t=2\pi$, connecting pairs of vertices as $K_2$ graphs for $t=2\pi$, or connecting four vertices as a $C_4$ and propagating for $t=\pi$. The best choice for implementation is likely to be determined by other scheduling concerns.
\par
\begin{figure}
\includegraphics[scale=.5]{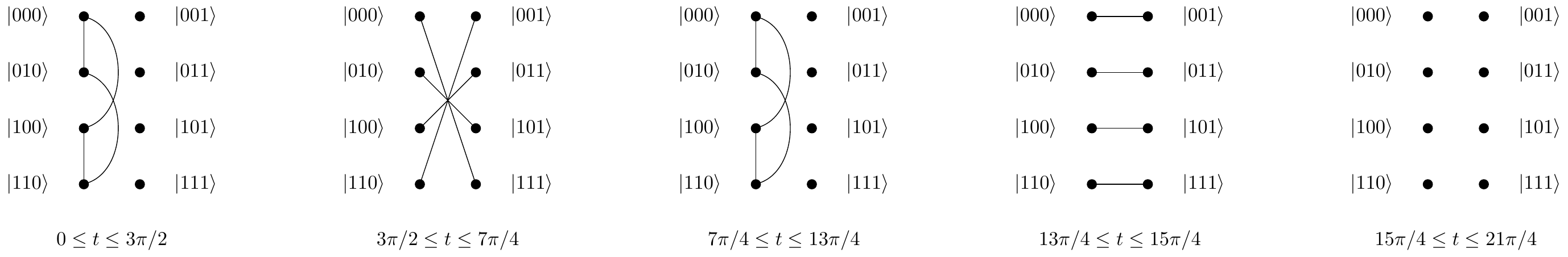}
\caption{A graphical representation of the $H$ gate using CTQW on $\mathcal{G}_{H}$.}
\label{fig:hadamardgate}
\end{figure}

The single-qubit Hadamard gate is defined in the computational basis as
\begin{equation}
H = \frac{1}{\sqrt{2}} \left(
\begin{array}{cc}
1 & 1 \\
1 & -1 
\end{array}
\right)
\end{equation}
and may be implemented using a series of ${C}_{4}$ and ${K}_{2}$ graphs. The Hadamard gate may be performed with only five vertices, but we again use eight vertices to establish a clear correspondence with three qubits in the circuit model. Consider the initial state $c_0 \ket{0} + c_1 \ket{1}$ embedded in a Hilbert space represented by $|V|=8$ nodes. Figure \ref{fig:hadamardgate} illustrates the dynamic graph for the $H$ gate, defined as 
\begin{equation}
\label{eq:h}
\begin{split}
\mathcal{G}_H = &\{(G_{C_4}^{(0,2,4,6)}+G_{K_1}^{(1)}+G_{K_1}^{(3)}+G_{K_1}^{(5)}+G_{K_1}^{(7)},3\pi/2),\\
&\quad (G_{K_2}^{(0,7)}+G_{K_2}^{(1,6)}+G_{K_2}^{(2,5)}+G_{K_2}^{(3,4)},\pi/4),\\
&\quad (G_{C_4}^{(0,2,4,6)}+G_{K_1}^{(1)}+G_{K_1}^{(3)}+G_{K_1}^{(5)}+G_{K_1}^{(7)},3\pi/2),\\ 
&\quad(G_{K_2}^{(0,1)}+G_{K_2}^{(2,3)}+G_{K_2}^{(4,5)}+G_{K_2}^{(6,7)},\pi/2),\\
&\quad (G_{K_1}^{(0)}+G_{K_1}^{(1)}+G_{K_1}^{(2)}+G_{K_1}^{(3)}+G_{K_1}^{(4)}+G_{K_1}^{(5)}+G_{K_1}^{(6)}+G_{K_1}^{(7)},3\pi/2) \}
\end{split}
\end{equation}
We show in the Appendix that the CTQW defined by Eq.~(\ref{eq:h}) implements the logical transformation for the Hadamard gate.
\par
\par
\begin{figure}
	\includegraphics[scale=.5]{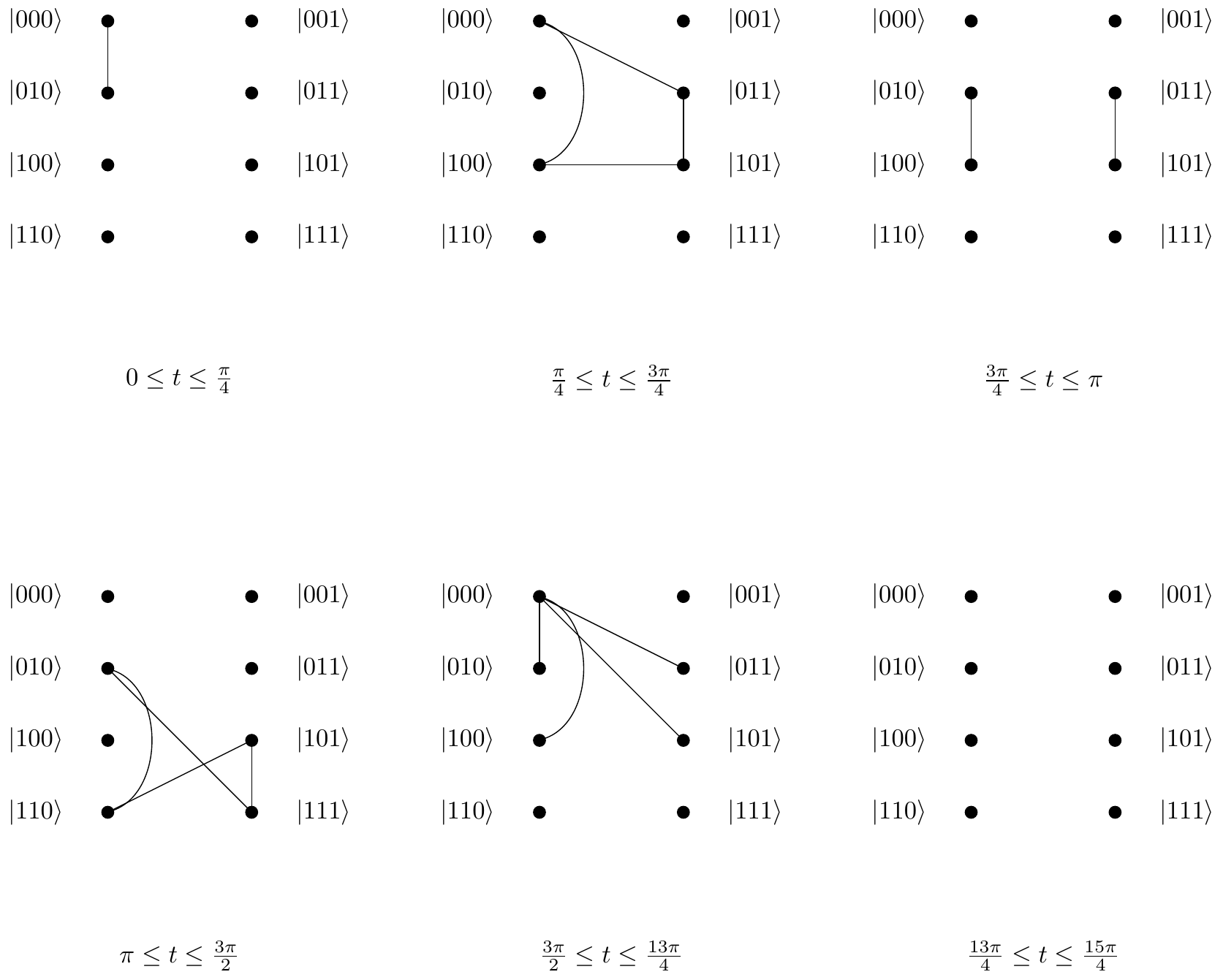}
	\caption{A graphical representation of the $T$ gate using $\mathcal{G}_{T}$.}
	\label{fig:tgate}
\end{figure}

The $T$ gate is defined as
\begin{equation}
T = \frac{1}{\sqrt{2}} \left(
\begin{array}{cc}
1 & 0 \\
0 & e^{\frac{i\pi}{4}} 
\end{array}
\right)
\end{equation}
and may be implemented using $K_1$, $K_2$, and $C_4$ graphs, along with the star graph on five vertices. A star graph is a connected graph $G$ on $n$ vertices such that exactly one vertex has degree $n-1$ and all other vertices have degree one. Figure \ref{fig:tgate} illustrates the dynamic graph used for the $T$ gate, which is written as
\begin{equation}
\begin{split}
\label{eq:t}
\mathcal{G}_{\textrm{T}} = 
&\{(G_{K_2}^{(0,2)} + G_{K_1}^{(1)}+G_{K_1}^{(3)} + G_{K_1}^{(4)}+G_{K_1}^{(5)} + G_{K_1}^{(6)}+G_{K_1}^{(7)}, \frac{\pi}{4}) , \\
& (G_{C_4}^{(0,3,4,5)}+G_{K_1}^{(1)} + G_{K_1}^{(2)} + G_{K_1}^{(6)} + G_{K_1}^{(7)}, \frac{\pi}{2}), \\
& (G_{K_2}^{(2,4)}+G_{K_2}^{(3,5)} + G_{K_1}^{(0)} + G_{K_1}^{(0)} + G_{K_1}^{(1)} + G_{K_1}^{(6)}+ G_{K_1}^{(7)}, \frac{\pi}{4}), \\
& (G_{C_4}^{(2,5,6,7)}+G_{K_1}^{(0)} + G_{K_1}^{(1)} + G_{K_1}^{(3)} + G_{K_1}^{(4)}, \frac{\pi}{2}) , \\
& (G_{S_5}^{(0,2,3,4,5)}+G_{K_1}^{(1)} + G_{K_1}^{(6)} + G_{K_1}^{(7)}, \frac{7\pi}{4}), \\
& (G_{K_1}^{(0)} + G_{K_1}^{(1)} + G_{K_1}^{(2)} + G_{K_1}^{(3)} + G_{K_1}^{(4)}+G_{K_1}^{(5)} + G_{K_1}^{(6)}+G_{K_1}^{(7)}, \frac{\pi}{4}) \}.
\end{split}
\end{equation}
We show in the Appendix that the CTQW defined by Eq.~(\ref{eq:t}) implements the logical transformation for the $T$ gate.
\par
\subsection{Multi-qubit gates}
\begin{figure}[h]
\includegraphics{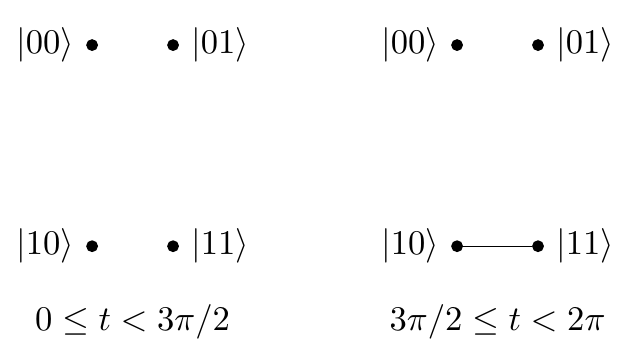}
\caption{A graphical representation of the CNOT gate using CTQW on $\mathcal{G}_{\textsc{cnot}}$.}
\label{fig:cxgate}
\end{figure}

Quantum walks on dynamics graphs may also be used to construct multi-qubits gates. For example, the two-qubit CNOT gate,
\begin{equation}
\textrm{CNOT} =\left(
\begin{array}{cccc}
1 & 0 & 0 & 0 \\
0 & 1 & 0 & 0 \\
0 & 0 & 0 & 1 \\
0 & 0 & 1 & 0 
\end{array}
\right)
\end{equation}
can be realized using a quantum walk on 4 vertices that span the space of the control and target qubits. Let vertices $0$ and $1$ propagate as singletons for time $2\pi$ while allowing vertices $2$ and $3$ to propagate under $K_2$ as shown in Fig.~\ref{fig:cxgate}.
\begin{equation}
\label{eq:cnot}
\mathcal{G}_{\textrm{CNOT}} = \{(G_{K_1}^{(0)}+G_{K_1}^{(1)} + G_{K_1}^{(2)}+G_{K_1}^{(3)}, \frac{3\pi}{2}) , (G_{K_1}^{(0)}+G_{K_1}^{(1)} + G_{K_2}^{(2,3)}, \frac{\pi}{2}) \}
\end{equation}
\par
\par
\begin{figure}[H]
\centering
\includegraphics{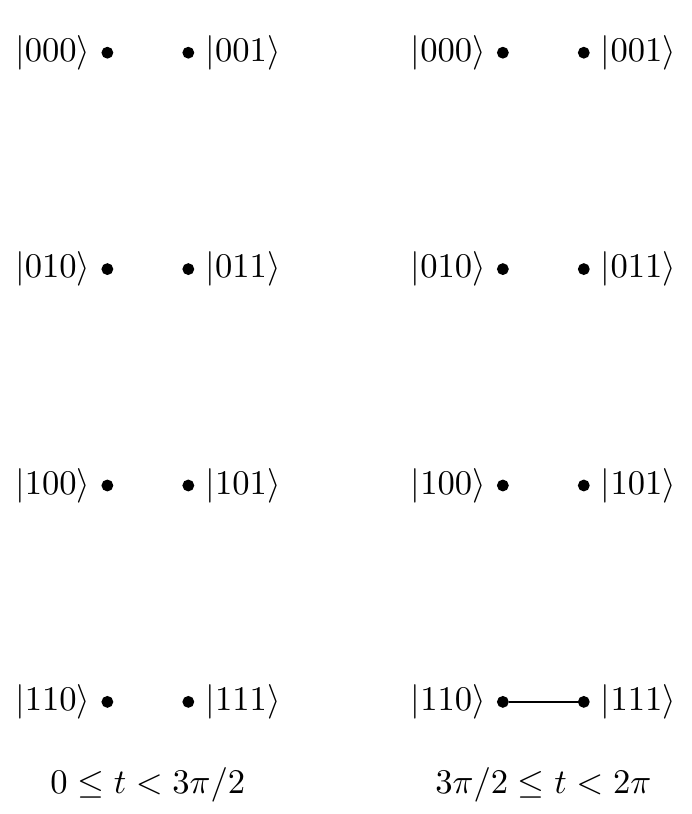}
\caption{A graphical representation of the CCNOT (Toffoli) gate using CTQW on $\mathcal{G}_{\textsc{CCNOT}}$.}
\label{fig:toffoligate}
\end{figure}

The three-qubit CCNOT, or Toffoli, gate is constructed similarly but now using $|V| = 8$ vertices that represent the two control qubits and one target qubit. The implementation of the Toffoli gate is identical to the $CNOT$ gate but with four additional vertices allowed to propagate as singletons for $2\pi$ units of time. It is used in both the carry and sum subcircuits in the quantum adder circuit. It is also reversible, meaning the its effects may be reversed using other operations. Figure \ref{fig:toffoligate} illustrates the dynamic graph for the Toffoli gate.
\subsection{Measurement and Initialization}
We model measurement of the quantum state on a graph $G$ as a projection onto a subspace of the basis $B_G$. In establishing a correspondence with the qubit-encoded circuit model, we decompose the labels of the basis according to a binary expansion
\begin{equation}
\ket{j} = \sum_{i=1}^{m}{j_{i} 2^{m-i}}
\end{equation}
with $j_{i} \in \{0,1\}$ and $m = \log_{2}|V|$. In this binary representation, a quantum state $\ket{\psi_{G}}\in B_{G}$ can be expressed as
\begin{equation}
\ket{\psi_{G}} = \sum_{j\in V}{{c_{j}\ket{j_{1},\ldots,j_{m}}}},
\end{equation}
and measuring the $i$-th qubit to have a fixed value $\bar{j}_i$ corresponds to projecting the state onto a subset of nodes in the graph, i.e.,
\begin{equation}
\ket{\bar{j}_{i}}\bracket{\bar{j}_i}{\psi_{G}} = \sum_{j\in V}{{c_{j}\ket{j_{1},\ldots,\bar{j}_i,\ldots,j_{m}}}}
\end{equation}
The probability to observe node $j$ is given as
\begin{equation}
\textrm{Prob}(j) =|\bracket{\bar{j}_{i}}{\psi_{G}}|^2= \sum_{j\in V, j_{i}=\bar{j}}{|c_{j}|^2 \leq 1}
\end{equation}
\par
We may use measurement as part of a deterministic initialization method, in which the projective outcome is transformed into the desired initial state. This requires conditional operations based on the decoded output from the measurement, from which the necessary series of single-qubit gates are applied to graph. For projections into the label basis, these feed-forward operations consist of products of the Pauli operators flip the label state to a fiducial starting label, e.g., the vertex 0.
\section{Quantum Walks From Quantum Circuits}
\label{sec:qwc}
We complete our analysis by providing explicit examples of how quantum walks on dynamic graphs realize circuits within gate-model computing. These examples highlight the differences in the representation of the logic as well as the resources required to achieve the desired unitary transformations. In our example, CTQWs are performed in series and the number of vertices needed to implement each circuit is equal to the largest of the number of vertices needed to perform the CTQW equivalent for each logic gate.
\subsection{Quantum Teleportation Circuit}
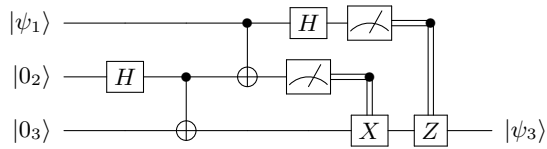
\begin{figure}[ht]
\centering
\mbox{
\Qcircuit @C=.7em @R=.4em @! {
\lstick{\ket{\psi_1}} 	&\qw			& \qw 		& \ctrl{1}	 	& \gate{H} 	& \meter 				& \control \cw \cwx[2] 	&  & \\
\lstick{\ket{0_2}} 		& \gate{H} 		& \ctrl{1} 	& \targ 		& \meter 	& \control \cw  \cwx[1]	& 						&   & \\
\lstick{\ket{0_3}} 		& \qw 			& \targ 		&\qw 		& \qw 		& \gate{X}  				& \gate{Z} 				&   \rstick{\ket{\psi_3}} \qw
}
}
\caption{The circuit model representation of quantum teleportation uses three qubits and a series of elementary gates.}
\label{fig:qtc}
\end{figure}

In quantum teleportation, a qubit of information is transferred from one logical element to another as shown in Fig.~\ref{fig:qtc}. In the circuit model description, three qubits are initially prepared in the state $\ket{000}$. The first element is prepared in the state $\ket{\psi_1}$ by applying the necessary single-qubit transformation. The remaining elements are prepared in a two-qubit entangled state by applying the Hadamard gate to the second element followed by the CNOT gate acting on the second and third elements. A second CNOT gate entangles the first and second qubits. A final Hadamard gate is applied to the first, after which measurements performed on elements 1 and 2 generate binary values $b_1$ and $b_2$, respectively. The effect of these measurements is to project element 3 into the state $X_3^{b_1}Z_{3}^{b_2} \ket{\psi_3}$, which may be transformed to the original state of element 1 with knowledge of $(b_1, b_2)$.
\par
\begin{figure}[ht]
\centering
\includegraphics[scale=0.4]{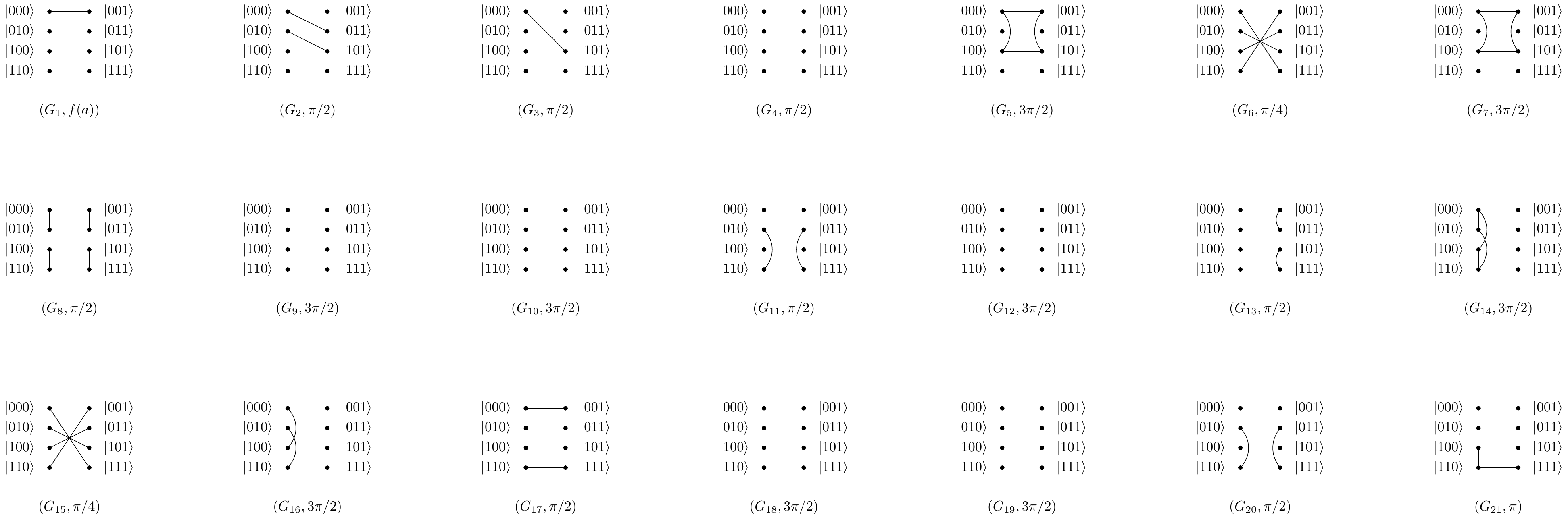}
\bigskip
\caption{In this graphical representation of quantum teleportation, each graph is labeled as $(G_{\ell}, \tau_{\ell})$ with $\tau_{\ell}$ the propagation time in the $\ell$-th graph. The time $f(a)=\textrm{arcsin}(\sqrt{a})$ is the state specific time required to rotate $\ket{000}$ to $\sqrt{1-a}\ket{000} +\sqrt{a}\ket{001}$. From left to right, the first four graphs rotate the state while the next five graphs correspond to the $H$ gate on the second qubit. The following four graphs represent a pair of CNOT gates. The next five graphs correspond with an $H$ gate on the first qubit. Assuming a measurement outcome $(b_1 = 1, b_2 = 1)$, the remaining graphs implement the $X$ and $Z$ gates needed to complete teleportation.}
\label{fig:qtc2}
\end{figure}

The implementation of quantum teleportation using CTQW on a dynamic graph is shown in Fig.~\ref{fig:qtc2}, and it begins with a graph on eight vertices. Initialization of these vertices is realized through a projective measurement and, depending on the measurement outcome, a sequence of $X$ operations to populate the 0 vertex. We then approximate an arbitrary unitary operation to prepare the input superposition state $\ket{\psi}=\sqrt{1-a}\ket{0} +\sqrt{a} \ket{1}$ for $ a\in \mathcal{C}$ where $|a|=1$. The number of vertices needed to represent an arbitrary $\ket{\psi}$ depends on the desired state, but this single-qubit unitary can be constructed using the universal basis described above. A Hadamard transform is then applied to vertices 0 and 7 using Eq.~(\ref{eq:h}) followed by a pair of CNOT transforms using Eq.~(\ref{eq:cnot}) acting on vertices $\{2,3,6,7 \}$ and $\{0, 1, 2, 3\}$, respectively. The output from this series of CTQWs prepares the graph state 
\begin{equation}
\ket{\psi} = \frac{1}{2} \left(-\sqrt{1-a} \ket{0} + \sqrt{1-a}\ket{1} +\sqrt{1-a} \ket{2} -\sqrt{1-a} \ket{3} -\sqrt{a} \ket{4} -\sqrt{a} \ket{5} + \sqrt{a}\ket{6} +\sqrt{a} \ket{7} \right)
\end{equation}
and a partial projective measurements on the first two bits of the label representation generates the four possible teleported states.

\begin{figure}[H]
\centering
\includegraphics[width=0.6\columnwidth]{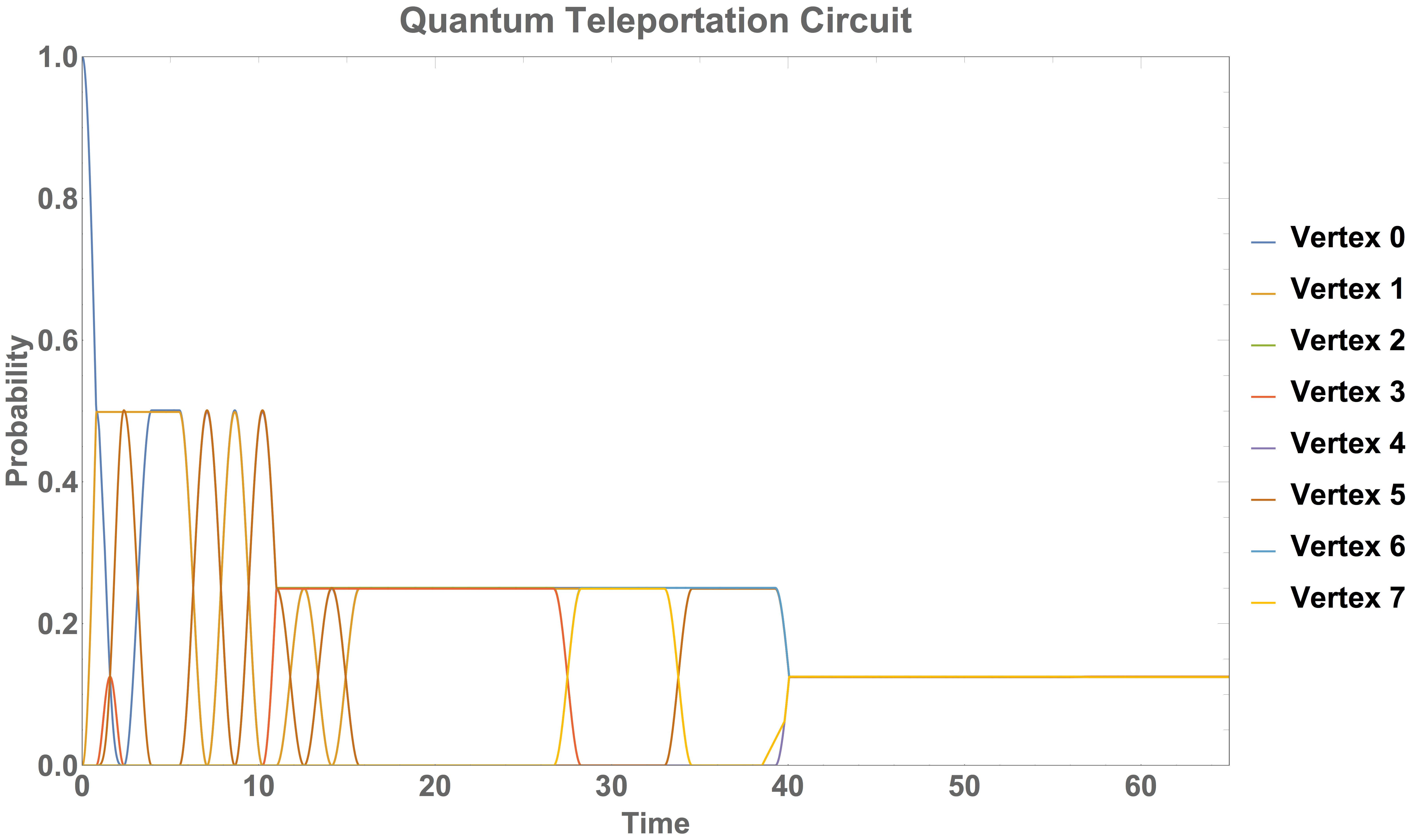}
\caption{The population dynamics for state preparation and quantum teleportation using CTQW on the dynamic graph shown in Fig.~\ref{fig:qtc2}.This examples corresponds to the case of measurement  outcomes $b_1 = 1$ and $b_2 = 1$ for qubits 1 and 2, respectively, and completes the protocol by applying the necessary recovery operations, $X$ and $Z$.}
\label{fig:qtcgraph}
\end{figure}

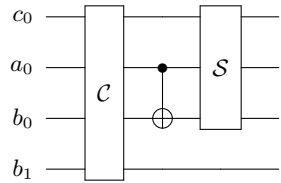
\begin{figure}[ht]
	\centering
	\mbox{
		\Qcircuit @C=.7em @R=.4em @! {
			\lstick{c_0} & \multigate{3}{\mathcal{C}} & \qw & \multigate{2}{\mathcal{S}}  & \qw\\
			\lstick{a_0} & \ghost{\mathcal{C}} & \ctrl{1}  & \ghost{\mathcal{S}}  & \qw \\
			\lstick{b_0} & \ghost{\mathcal{C}} & \targ  & \ghost{\mathcal{S}} & \qw \\
			\lstick{b_1} & \ghost{\mathcal{C}} & \qw  & \qw & \qw\\
		}
	}
\caption{A quantum circuit for addition of two 1-bit numbers, where the carry circuit $\mathcal{C}$ and the sum circuit $\mathcal{S}$ are defined in Figs.~\ref{fig:carry} and ~\ref{fig:sum}. Note that since we only have one carry operation, it is our last carry, and thus is not reversed.}
\label{fig:qa}
\end{figure}
\subsection{Quantum Adder}
As a second example, we consider a quantum addition circuit for summing two positive integers such that the input $\ket{a,b} \rightarrow \ket{a,a\oplus b}$ \cite{Vedral1996}. This variant of in-place addition takes two inputs encoded in registers $a$ and $b$ with the binary representations $a=a_{n-1} a_{n-2} ... a_1 a_0$ and $b = b_{n-1} b_{n-2} ... b_1 b_0$. 
An additional bit $b_{n+1} = 0$ is added to register $b$ to give a size $n+1$. 
A third workspace register $c$ of size $n-1$ is used in this implementation to store carry values with initialization $c_i = 0 \; \forall \; i$, while the final carry value is stored in the bit $b_{n+1}$. The circuit is composed from two subcircuits for carry and sum operations denoted as $\mathcal{C}$ and $\mathcal{S}$, respectively, and the subcircuits for $\mathcal{C}$ and $\mathcal{S}$ are specified in Figs.~\ref{fig:carry} and ~\ref{fig:sum}, respectively. The carry operation uses a Toffoli gate with the second and third qubit as controls and the fourth qubit as the target. This is followed by a CNOT gate on the second and third qubits before another Toffoli gate on the first, second, and fourth qubits. The reverse carry $\mathcal{RC}$ circuit undoes the carry computation by applying the gates in the reverse order. The last carry bit in the computation is not reversed but stored as $b_{n+1}$. The sum subcircuit denoted as $\mathcal{S}$ in Fig.~\ref{fig:sum} takes three qubits as input, in which a CNOT is applied to the second and third qubits followed by a Toffoli gate performed with the first two qubits being the controls and the third qubit as the target. In Fig.~\ref{fig:qa}, we show the demonstrated instance of one-bit inputs, i.e., $n = 1$, for which the reverse carry subcircuit is unnecessary. For this example, carry bits are also uncessary but we include the single carry bit $c_0$ to confirm generality.
\begin{figure}[H]
	\centering
	\mbox{
\Qcircuit @C=.7em @R=.4em @! {
& \qw & \qw & \ctrl{2} & \qw \\
& \ctrl{1} & \ctrl{1} & \qw & \qw \\
& \ctrl{1} & \targ & \ctrl{1} & \qw \\
& \targ & \qw & \targ & \qw
}
}
\caption{The carry subcircuit $\mathcal{C}$ used in Fig.~\ref{fig:qa}}
\label{fig:carry}
\end{figure}
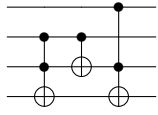
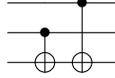
\begin{figure}[H]
	\centering
	\mbox{
\Qcircuit @C=.7em @R=.4em @! {
& \qw & \ctrl{2} & \qw\\
& \ctrl{1} & \qw & \qw\\
& \targ & \targ& \qw
}
}
\caption{The sum subcircuit $\mathcal{S}$ used in Fig.~\ref{fig:qa}}
\label{fig:sum}
\end{figure}
\par
\begin{figure}[H]
	\centering
	\includegraphics[scale=0.35]{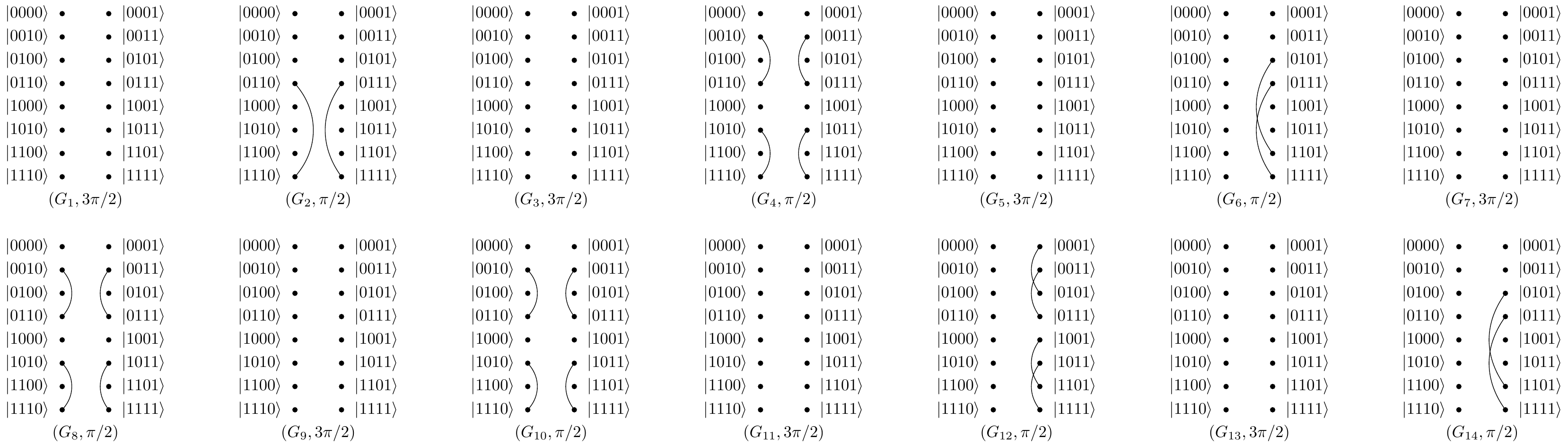}
	\caption{In this graphical representation of a one-bit quantum adder circuit, each graph is labeled as $(G_{\ell}, \tau_{\ell})$ with $\tau_{\ell}$ the propagation time in the $\ell$-th graph.}
	\label{fig:qa2}
\end{figure}

We reduce the gate sequences in the quantum addition circuit into the dynamic graph shown in Fig.~\ref{fig:qa2}. Our reduction uses the CTQWs for CNOT and CCNOT gates described in Sec.~\ref{sec:qweg} and sequentially orders them according to the gate specification in Figs.~\ref{fig:qa},~\ref{fig:carry}, and ~\ref{fig:sum}. In order to verify the correctness of the reduction, we have used numerical simulation to determine the quantum state generated by the CTQW on the dynamic graph shown in Fig.~\ref{fig:qa2}. Numerical simulation of the CTQW requires a memory space that is exponential in the number of qubits, i.e, $2^{3n+1}$
. Implementing the quantum adder circuit for $n=1$ requires a dynamic graph on sixteen vertices. 
\par
\begin{figure}[H]
	\centering
	\includegraphics[width=0.8\columnwidth]{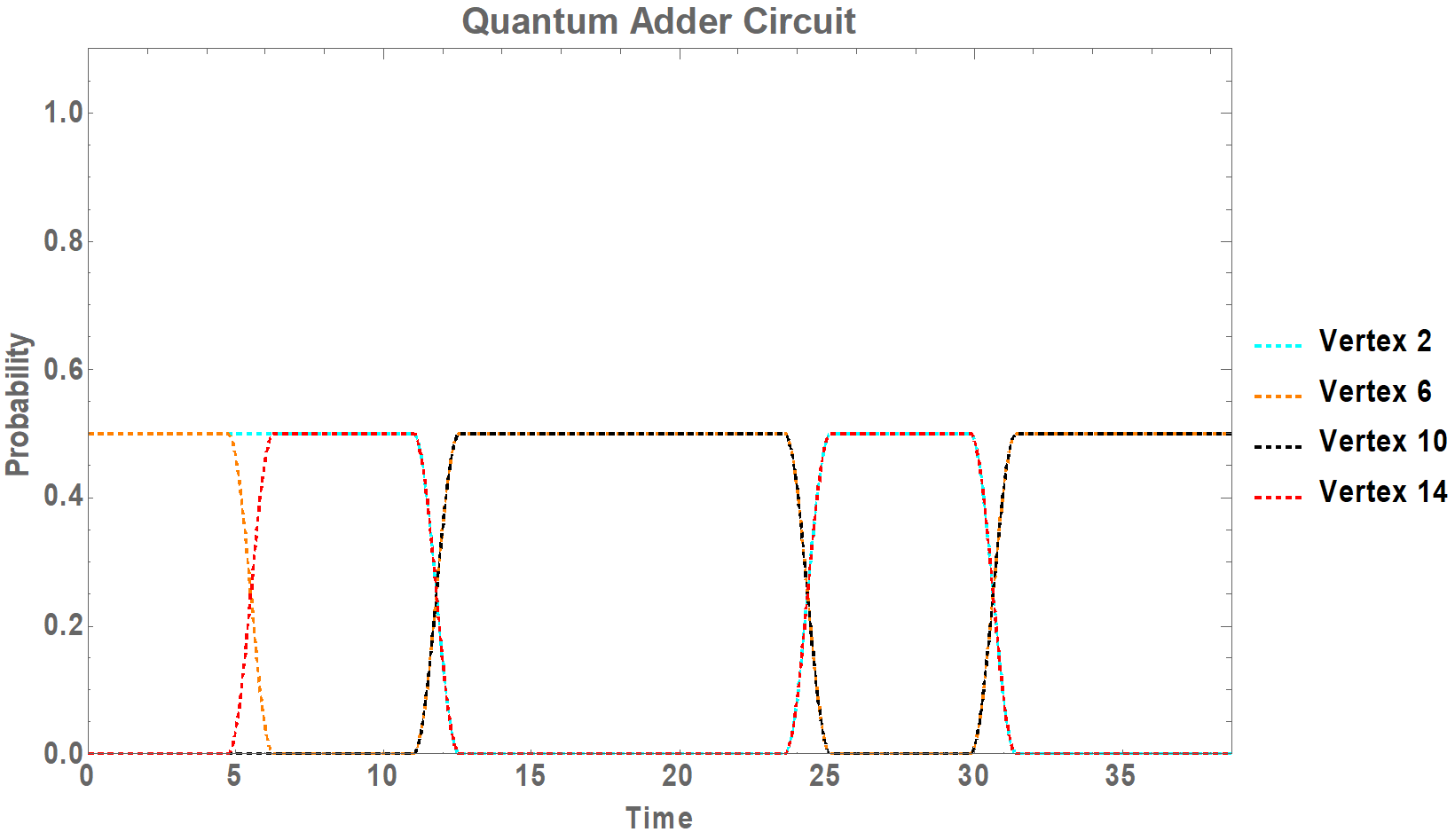}
	\caption{The population dynamics of the CTQW for quantum addition of inputs $\ket{a}=\ket{1}$ and $\ket{b}=\frac{1}{\sqrt{2}}\ket{0,0} + \frac{1}{\sqrt{2}}\ket{0,1}$. Numerical simulations of the CTQW on the dynamic graph shown in Fig.~\ref{fig:qa2} calculates exactly the amplitudes of each vertex and the final state is $\ket{b_1, b_0,a_0,c_0} = \frac{1}{\sqrt{2}}(\ket{0,1,1,0} + \ket{1,0,1,0})$, which corresponds to a uniform superposition of the vertices $6$ and $10$.}
	\label{fig:qaplotmixed}
	\end{figure}

We show results from a specific simulation with $\ket{a_0} =  \ket{1}$ and $\ket{b_0} = \frac{1}{\sqrt{2}} \left(\ket{0} + \ket{1}\right)$ in Fig.~\ref{fig:qaplotmixed}. We plot the time-dependent population of the vertices that represent the joint state of the computational registers. The carry register is initialized to $\ket{c_0} = \ket{0}$ and the resulting computational output is $\ket{b_1, b_0,a_0,c_0} = \frac{1}{\sqrt{2}}\left(\ket{0,0,1,0} + \ket{1,1,1,0}\right)$,  where the $a_0$ and $c_0$ registers remain in their initial states, and the sum $a_0 + b_0$ is stored in the $b_0$ and $b_1$. As shown in Fig.~\ref{fig:qaplotmixed}, our CTQW simulations verify that the dynamic graph yields the expected output states, which corresponds to a uniform superposition of the vertex labels $6$ and $10$.


\section{Discussion}
\label{sec:con}
Continuous-time quantum walks offer a versatile paradigm for quantum computing, in which the edges between vertices in a graph serve to model the connectivity between basis states. We have defined a dynamic graph as a time-ordered sequence of changing connectivity through which a the state of a continuous quantum walk can be tailored to perform computation and, in particular, we have provided constructions of continuous-time quantum walks on dynamic graphs that implement a diverse set of gates taken from the quantum circuit computational model. Our realizations of the single-qubit Pauli, Hadamard, and $T$ gates, and the CNOT and Toffoli gates, as well as measurement and initialization, form a complete set of primitive operations that can be composed to approximate an arbitrary unitary operator. We have presented implementations of the the bit-wise addition operation and quantum teleportation to demonstrate composition of quantum walks and shown how some reduction in the composite dynamic graph can be realized by eliminating redundancies.
\par 
An important distinction in our formulation of continuous-time quantum walks is the condition that the Hamiltonian represent the connectivity of the underlying basis states. Although we permit variations in this Hamiltonian, the restriction has several side-effects on the computational model. For example, our design for some single-qubit gates taken from the circuit model require graphs with more than two vertices. These additional vertices are effectively ancilla used to store temporarily intermediate states of the walk. This unique representation may afford opportunities for optimizing quantum logic by better understanding the transformation of an input state to its output form. Similarly, multi-qubit gates such as CNOT and Toffoli are trivial to implement by using the starkly different periods for perfect state transfer. Algorithmic methods that take advantage of these otherwise idle vertices may provide more compact representations of logical transformations.
\par
We have restricted designs of the current quantum walks to small and relatively simple graphs, e.g, $K_2$ and $C_4$. These designs are appealing because they require less complex interactions between the physical elements, but the ability to realize these designs will depend on technological constraints as well as algorithmic requirements. In particular, perfect state transfer has been implemented recently in a photonic processor \cite{chapman2016experimental}. Chapman et al.~used a linear array of evanescently coupled waveguides to realize nearest-neighbor coupling and transfer the polarized state of one photon to another. The underlying tight-binding Hamiltonian provides an approximation to the connectivity graph underlying a continuous-time quantum walk defined within the space of the single-photon Fock states. The approximation is controlled by the spectra of the coupled waveguides, which must be non-uniform in their geometry for (almost) perfect state transfer using a linear coupled chain \cite{Christandl2004,Christandl2005}. The geometrical constraints imposed by linear chains have been overcome by a recent demonstration of continuous-time quantum walks in two-dimensional waveguide arrays \cite{tang2018experimental}. Tang et al.~demonstrated control of the coupling between waveguide in a two-dimensional array by fabricating specific distance between the channels. We anticipate that these capabilities may be applied to vary the coupling along the waveguide length and, consequently, develop a physical realization of a dynamic graph. These adaptations may require relaxations of our model, including modifying the sharp transitions induced the rectangle function with more gradual transitions.
\section*{Acknowledgments}
This work was supported in part by the U.S. Department of Energy, Office of Science, Office of Workforce Development for Teachers and Scientists (WDTS) under the Science Undergraduate Laboratory Internships Program (SULI) as well as the Department of Energy, Office of Science Early Career Research Program and the Mathematical Sciences Graduate Internship program of the National Science Foundation. 

\section*{Appendix}
We demonstrate that the dynamic graph representing Eq.~(\ref{eq:h}) implements the Hadamard transform by showing explicitly the graph state prepared under the sequence of CTQWs. We first note that the CTQW on each element $G_{\ell}$ in a dynamic graph can be evaluated numerically for the designated propagation time $t_{\ell}$. For $\mathcal{G}_{H}$, we have
\begin{align*}
U_{G_{0}} &= \left(
\begin{array}{cccccccc}
0 & 0 & 0 & 0 & 0 & 0 & -1 & 0 \\
0 & i & 0 & 0 & 0 & 0 & 0 & 0 \\
0 & 0 & 0 & 0 & -1 & 0 & 0 & 0 \\
0 & 0 & 0 & i & 0 & 0 & 0 & 0 \\
0 & 0 & -1 & 0 & 0 & 0 & 0 & 0 \\
0 & 0 & 0 & 0 & 0 & i & 0 & 0 \\
-1 & 0 & 0 & 0 & 0 & 0 & 0 & 0 \\
0 & 0 & 0 & 0 & 0 & 0 & 0 & i \\
\end{array}
\right)
\end{align*}
\begin{align*}
U_{G_{1}} &= \left(
\begin{array}{cccccccc}
\frac{1}{\sqrt{2}} & 0 & 0 & 0 & 0 & 0 & 0 & \frac{-i}{\sqrt{2}} \\
0 & \frac{1}{\sqrt{2}} & 0 & 0 & 0 & 0 & \frac{-i}{\sqrt{2}} & 0 \\
0 & 0 & \frac{1}{\sqrt{2}} & 0 & 0 & \frac{-i}{\sqrt{2}} & 0 & 0 \\
0 & 0 & 0 & \frac{1}{\sqrt{2}} & \frac{-i}{\sqrt{2}} & 0 & 0 & 0 \\
0 & 0 & 0 & \frac{-i}{\sqrt{2}} & \frac{1}{\sqrt{2}} & 0 & 0 & 0 \\
0 & 0 & \frac{-i}{\sqrt{2}} & 0 & 0 & \frac{1}{\sqrt{2}} & 0 & 0 \\
0 & \frac{-i}{\sqrt{2}} & 0 & 0 & 0 & 0 & \frac{1}{\sqrt{2}} & 0 \\
\frac{-i}{\sqrt{2}} & 0 & 0 & 0 & 0 & 0 & 0 & \frac{1}{\sqrt{2}} \\
\end{array}
\right)
\end{align*}
\begin{align*}
U_{G_{2}} &= \left(
\begin{array}{cccccccc}
0 & 0 & 0 & 0 & 0 & 0 & -1 & 0 \\
0 & -i & 0 & 0 & 0 & 0 & 0 & 0 \\
0 & 0 & 0 & 0 & -1 & 0 & 0 & 0 \\
0 & 0 & 0 & -i & 0 & 0 & 0 & 0 \\
0 & 0 & -1 & 0 & 0 & 0 & 0 & 0 \\
0 & 0 & 0 & 0 & 0 & -i & 0 & 0 \\
0 & 0 & 0 & 0 & 0 & 0 & 0 & -1 \\
-i & 0 & 0 & 0 & 0 & 0 & 0 & 0 \\
\end{array}
\right) \\
\end{align*}

\begin{align*}
U_{G_{3}} &= \left(
\begin{array}{cccccccc}
0 & -i & 0 & 0 & 0 & 0 & 0 & 0 \\
-i & 0 & 0 & 0 & 0 & 0 & 0 & 0 \\
0 & 0 & 0 & -i & 0 & 0 & 0 & 0 \\
0 & 0 & -i & 0 & 0 & 0 & 0 & 0 \\
0 & 0 & 0 & 0 & 0 & -i & 0 & 0 \\
0 & 0 & 0 & 0 & -i & 0 & 0 & 0 \\
0 & 0 & 0 & 0 & 0 & 0 & 0 & -i \\
0 & 0 & 0 & 0 & 0 & 0 & -i & 0 \\
\end{array}
\right) \\
\end{align*}

\begin{align*}
U_{G_{4}} &= \left(
\begin{array}{cccccccc}
i & 0 & 0 & 0 & 0 & 0 & 0 & 0 \\
0 & i & 0 & 0 & 0 & 0 & 0 & 0 \\
0 & 0 & i & 0 & 0 & 0 & 0 & 0 \\
0 & 0 & 0 & i & 0 & 0 & 0 & 0 \\
0 & 0 & 0 & 0 & i & 0 & 0 & 0 \\
0 & 0 & 0 & 0 & 0 & i & 0 & 0 \\
0 & 0 & 0 & 0 & 0 & 0 & i & 0 \\
0 & 0 & 0 & 0 & 0 & 0 & 0 & i \\
\end{array}
\right) 
\end{align*}
By multiplying the resulting matrices in order, we construct an explicit numerical representation for the CTQW under the dynamic graph $\mathcal{G}_H$ as
\begin{equation*}
U_{\mathcal{G}_H} = 
\left( 
\begin{array}{cccccccc}
\frac{1}{\sqrt{2}} & \frac{1}{\sqrt{2}} & 0 & 0 & 0 & 0 & 0 & 0 \\
\frac{1}{\sqrt{2}} & \frac{-1}{\sqrt{2}} & 0 & 0 & 0 & 0 & 0 & 0 \\
0 & 0 & \frac{1}{\sqrt{2}} & \frac{1}{\sqrt{2}} & 0 & 0 & 0 & 0 \\
0 & 0 & \frac{1}{\sqrt{2}} & \frac{-1}{\sqrt{2}} & 0 & 0 & 0 & 0 \\
0 & 0 & 0 & 0 & \frac{1}{\sqrt{2}} & \frac{1}{\sqrt{2}} & 0 & 0 \\
0 & 0 & 0 & 0 & \frac{1}{\sqrt{2}} & \frac{-1}{\sqrt{2}} & 0 & 0 \\
0 & 0 & 0 & 0 & 0 & 0 & \frac{1}{\sqrt{2}} & \frac{1}{\sqrt{2}} \\
0 & 0 & 0 & 0 & 0 & 0 & \frac{1}{\sqrt{2}} & \frac{-1}{\sqrt{2}}
\end{array}
\right) 
\end{equation*}
It is then apparent from this numerical representation that the CTQW for $\mathcal{G}_{H}$ is equivalent to applying the circuit-model operator $H_{1} \otimes H_{2} \otimes H_{3}$ on the three-qubit Hilbert space.
\par
We provide a similar proof that the dynamic graph representing Eq.~(\ref{eq:t}) implements the $T$ gate by showing explicitly the graph state prepared under the sequence of CTQWs. We first note that
\begin{align*}
U_{G_{0}} &= \left(
\begin{array}{cccccccc}
\frac{1}{\sqrt{2}} & 0 & \frac{-i}{\sqrt{2}} & 0 & 0 & 0 & 0 & 0 \\
0 & e^{\frac{-i\pi}{4}} & 0 & 0 & 0 & 0 & 0 & 0 \\
\frac{-i}{\sqrt{2}} & 0 & \frac{1}{\sqrt{2}} & 0 & 0 & 0 & 0 & 0 \\
0 & 0 & 0 &  e^{\frac{-i\pi}{4}} & 0 & 0 & 0 & 0 \\
0 & 0 & 0 & 0 &  e^{\frac{-i\pi}{4}} & 0 & 0 & 0 \\
0 & 0 & 0 & 0 & 0 &  e^{\frac{-i\pi}{4}} & 0 & 0 \\
0 & 0 & 0 & 0 & 0 & 0 &  e^{\frac{-i\pi}{4}} & 0 \\
0 & 0 & 0 & 0 & 0 & 0 & 0 &  e^{\frac{-i\pi}{4}} \\
\end{array}
\right)
\end{align*}

\begin{align*}
U_{G_{1}} &= \left(
\begin{array}{cccccccc}
0 & 0 & 0 & 0 & 0 & -1 & 0 & 0 \\
0 & -i & 0 & 0 & 0 & 0 & 0 & 0 \\
0 & 0 & -i & 0 & 0 & 0 & 0 & 0 \\
0 & 0 & 0 & 0 & -1 & 0 & 0 & 0 \\
0 & 0 & 0 & -1 & 0 & 0 & 0 & 0 \\
-1 & 0 & 0 & 0 & 0 & 0 & 0 & 0 \\
0 & 0 & 0 & 0 & 0 & 0 & -i & 0 \\
0 & 0 & 0 & 0 & 0 & 0 & 0 & -i \\
\end{array}
\right)
\end{align*}

\begin{align*}
U_{G_{2}} &= \left(
\begin{array}{cccccccc}
e^{\frac{-i\pi}{4}} & 0 & 0 & 0 & 0 & 0 & 0 & 0 \\
0 & e^{\frac{-i\pi}{4}} & 0 & 0 & 0 & 0 & 0 & 0 \\
0 & 0 & \frac{1}{\sqrt{2}} & 0 & \frac{-i}{\sqrt{2}} & 0 & 0 & 0 \\
0 & 0 & 0 & \frac{1}{\sqrt{2}} & 0 & \frac{-i}{\sqrt{2}} & 0 & 0 \\
0 & 0 & \frac{-i}{\sqrt{2}} & 0 & \frac{1}{\sqrt{2}} & 0 & 0 & 0 \\
0 & 0 & 0 & \frac{-i}{\sqrt{2}} & 0 & \frac{1}{\sqrt{2}} & 0 & 0 \\
0 & 0 & 0 & 0 & 0 & 0 & e^{\frac{-i\pi}{4}} & 0 \\
0 & 0 & 0 & 0 & 0 & 0 & 0 & e^{\frac{-i\pi}{4}} \\
\end{array}
\right)
\end{align*}

\begin{align*}
U_{G_{3}} &= \left(
\begin{array}{cccccccc}
-i & 0 & 0 & 0 & 0 & 0 & 0 & 0 \\
0 & -i & 0 & 0 & 0 & 0 & 0 & 0 \\
0 & 0 & 0 & 0 & 0 & -1 & 0 & 0 \\
0 & 0 & 0 & -i & 0 & 0 & 0 & 0 \\
0 & 0 & 0 & 0 & -i & 0 & 0 & 0 \\
0 & 0 & -1 & 0 & 0 & 0 & 0 & 0 \\
0 & 0 & 0 & 0 & 0 & 0 & 0 & -1 \\
0 & 0 & 0 & 0 & 0 & 0 & -1 & 0 \\
\end{array}
\right)
\end{align*}

\begin{align*}
U_{G_{4}} &= \left(
\begin{array}{cccccccc}
0 & 0 & \frac{i}{2} & \frac{i}{2} & \frac{i}{2} & \frac{i}{2} & 0 & 0 \\
0 & e^{\frac{i\pi}{4}} & 0 & 0 & 0 & 0 & 0 & 0 \\
\frac{i}{2} & 0 & \frac{3}{4} & \frac{-1}{4} & \frac{-1}{4} & \frac{-1}{4} & 0 & 0 \\
\frac{i}{2} & 0 & \frac{-1}{4} & \frac{3}{4} & \frac{-1}{4} & \frac{-1}{4} & 0 & 0 \\
\frac{i}{2} & 0 & \frac{-1}{4} & \frac{-1}{4} & \frac{3}{4} & \frac{-1}{4} & 0 & 0 \\
\frac{i}{2} & 0 & \frac{-1}{4} & \frac{-1}{4} & \frac{-1}{4} & \frac{3}{4} & 0 & 0 \\
0 & 0 & 0 & 0 & 0 & 0 & e^{\frac{i\pi}{4}} & 0 \\
0 & 0 & 0 & 0 & 0 & 0 & 0 & e^{\frac{i\pi}{4}} \\
\end{array}
\right) 
\end{align*}

\begin{align*}
U_{G_{5}} &= \left(
\begin{array}{cccccccc}
-i & 0 & 0 & 0 & 0 & 0 & 0 & 0 \\
0 & -i & 0 & 0 & 0 & 0 & 0 & 0 \\
0 & 0 & -i & 0 & 0 & 0 & 0 & 0 \\
0 & 0 & 0 & -i & 0 & 0 & 0 & 0 \\
0 & 0 & 0 & 0 & -i & 0 & 0 & 0 \\
0 & 0 & 0 & 0 & 0 & -i & 0 & 0 \\
0 & 0 & 0 & 0 & 0 & 0 & -i & 0 \\
0 & 0 & 0 & 0 & 0 & 0 & 0 & -i \\
\end{array}
\right) 
\end{align*}
Thus, as $ \mathcal{G}_T$ is the product of the above matrices, we have that 

\begin{equation*}
U_{\mathcal{G}_T} = \left( \begin{array}{cccccccc}
1 & 0 & 0 & 0 & 0 & 0 & 0 & 0 \\
0 & e^{\frac{i\pi}{4}} & 0 & 0 & 0 & 0 & 0 & 0 \\
0 & 0 & \frac{-1}{2} & 0 & \frac{-e^{\frac{-i\pi}{4}}}{\sqrt{2}} & \frac{1}{2} & 0 & 0 \\
0 & 0 & \frac{-1}{2} & 0 & \frac{e^{\frac{-i\pi}{4}}}{\sqrt{2}} & \frac{1}{2} & 0 & 0 \\
0 & 0 & \frac{1}{2} & \frac{e^{\frac{-i\pi}{4}}}{\sqrt{2}} & 0 & \frac{1}{2} & 0 & 0 \\
0 & 0 & \frac{1}{2} & \frac{-e^{\frac{-i\pi}{4}}}{\sqrt{2}} & 0 & \frac{1}{2} & 0 & 0 \\
0 & 0 & 0 & 0 & 0 & 0 & 0 & e^{-\frac{i\pi}{4}} \\
0 & 0 & 0 & 0 & 0 & 0 & e^{-\frac{i\pi}{4}} & 0 \\
\end{array}
\right) 
\end{equation*}

\bibliographystyle{apsrev4-1}
\bibliography{rr.manuscript}

\end{document}